# Truthful Feedback for Sanctioning Reputation Mechanisms

**Jens Witkowski**
Department of Computer Science
Albert-Ludwigs-Universität
Freiburg, Germany
witkowsk@informatik.uni-freiburg.de

## Abstract

For product rating environments, similar to that of Amazon Reviews, it has been shown that the truthful elicitation of feedback is possible through mechanisms which pay buyer reports contingent on the reports of other buyers. We study whether similar mechanisms can be designed for reputation mechanisms at online auction sites where the buyers' experiences are partially determined by a strategic seller. We show that this is impossible for the basic setting. However, introducing a small prior belief that the seller is a cooperative commitment player leads to a payment scheme with a truthful perfect Bayesian equilibrium.

## 1 INTRODUCTION

Virtually every type of good is traded online: books and electronic devices are ordered from Amazon[1], used items are traded on eBay[2] and services are offered at freelance portals, such as Elance[3]. Almost every e-commerce site employs a so-called reputation mechanism which collects and publishes ratings from their customers and it is instructive to distinguish between different kinds of reputation mechanisms in accordance with the role they play: those that are employed by online opinion forums, such as Amazon Reviews or Ciao[4], are built to eliminate asymmetric information. They are called *signaling* mechanisms as they collect and publish feedback about the quality of products and thus signal the previously hidden quality to future customers. This is different for the reputation mechanisms employed at online auction sites, such as eBay. These mechanisms are called *sanctioning* as they induce cooperation of the auction sellers through a threat to sanction uncooperative behavior. The essential distinction between these two roles of reputation mechanisms is the type of seller behavior: in signaling settings, sellers differ in their *abilities*. That is, some sellers are of higher quality than others. In sanctioning settings, all sellers are equally able but *have a temptation to cheat* as cooperation involves higher costs.

A common feature of almost all reputation mechanisms is the dependency on honest feedback. Most mechanisms in the literature simply *assume* that feedback is reported honestly. From a game-theoretic point of view, this assumption is problematic for two reasons: the first is the customers' motivation to participate at all. The feedback procedure requires a buyer to register an account, to log in and to fill out forms describing her[5] experiences. While this is time consuming and thus costly, the reported information benefits other customers but not the posting customer herself, so that standard economic theory predicts an under-provision of feedback. The second difficulty is honesty. External interests, i. e. biases towards dishonest reporting, may come from a variety of motivations. As an example for online auction sites, imagine two sellers competing for the same group of customers. Either seller has an incentive to badmouth its competitor, to praise his own service or to pay a customer to do so. A particularly common issue at online auction sites with bi-directional feedback, i. e. where both the seller and the buyer can rate the transaction, is retaliatory feedback. Empirical data of eBay's pre-2008 reputation mechanism suggests that sellers wait until the buyers have posted their feedback and match it thereafter. That is, sellers post the same rating they have received from the buyers and, in particular, retaliate against negative feedback (Resnick and Zeckhauser 2002).

---

[1] www.amazon.com
[2] www.ebay.com
[3] www.elance.com
[4] www.ciao.com

---

[5] We refer to buyers and sellers as female and male, respectively.

The truthful elicitation of feedback is thus crucial to incorporate into the design of sanctioning reputation mechanisms. This task is difficult because both the sellers' actions and the subsequent outcomes are private information which are never publicly revealed. One way to induce honest feedback in *signaling* settings is to pay a buyer for her feedback report conditional on the report of another buyer. This so-called "peer prediction method" was introduced by Miller, Resnick and Zeckhauser (2005) and considerably extended by Jurca and Faltings (2006; 2009). With regard to truthful feedback elicitation, signaling mechanisms are convenient because the perceived quality is essentially identical for all customers. Consider a digital camera bought via Amazon as an example: while different customers may experience different quality levels due to noise, they all receive the identical model. This is different for the customers' signals at eBay as they primarily depend on the sellers' *actions*, i.e. whether or not the respective seller sent the good in the prescribed quality. To our knowledge, the only *sanctioning* reputation mechanism that incorporates truthful feedback is the CONFESS mechanism developed by Jurca and Faltings (2007). Unfortunately, as it resorts to the folk theorems to induce cooperation, it inherits their dependency on repeated interactions which rules out its application at online auction sites.

In this paper, we study the mechanism design space of peer-based feedback elicitation for eBay-like online auctions. In the game-theoretic literature on sanctioning reputation mechanisms it is usually assumed that feedback is honest and one then studies how the degree of seller cooperation can be increased by tweaking the design parameters. For example, Dellarocas (2005) studies how the length of the feedback history influences seller cooperation. In this paper, *we take the complementary view* and assume that we are given a sanctioning mechanism which induces a certain degree of seller cooperation under the assumption of honest feedback. That is, for those settings for which we retrieve positive results, the feedback mechanisms which we provide can be merged with a sanctioning mechanism, such as the one developed by Dellarocas, into a fully incentivized sanctioning reputation mechanism.

# 2 THE MODEL

Before we study the design space of feedback mechanisms for online auction sites, we explain the basic procedure together with the game-theoretic model.

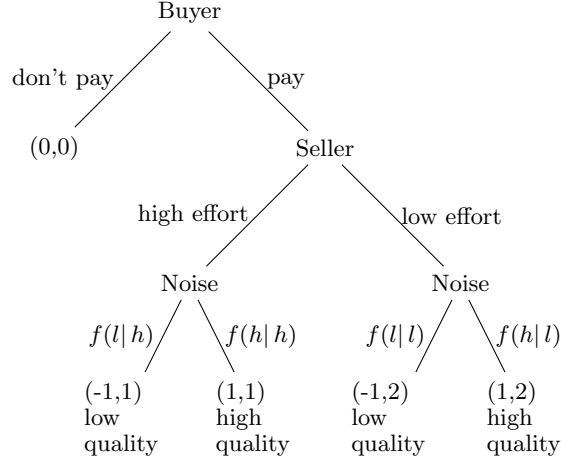

Figure 1: The game tree for a single online auction game with imperfect monitoring and noise parameter $f(\cdot|\cdot)$ with $0 < f(h|l) < f(h|h) < 1$. Note that the given payoffs are example numbers with the first and the second number denoting the buyer's and the seller's payoff, respectively.

## 2.1 PROCEDURE

The idealized procedure of a single transaction is the following: the seller describes the quality of a good that he wants to sell and interested customers submit their bids. At some point, the auction site determines the buyer who subsequently pays for the good. After reception of payment, the seller sends it in the prescribed quality. Unfortunately, this idealized procedure is prone to cheating: since the buyer of the good has to pay for it before the seller sends it, the seller is better off by cheating, i.e. either not sending the good at all or in a quality that is lower than described. Anticipating this, the buyer does not pay for the good in the first place. This is called a dilemma as it results in a no-trade situation even if both buyer and seller gained from trade if the seller could credibly commit to cooperation. In the following, we restrict ourselves to settings with numerous effort levels but only binary signals. The interpretation of a signal is whether the contract, i.e. the good's description together with the promise of sending it, was fulfilled. The procedure with imperfect monitoring is depicted in Figure 1.

## 2.2 SELLER BEHAVIOR

We make the usual assumption for sanctioning mechanisms that the seller is a long-run player who faces a sequence of short-run buyers, i.e. buyers who are interested in the seller's product in only a single stage game. Every buyer plays a best response to the belief of the seller's action. That is, a buyer decides to pay for the good if the belief that the seller will cooperate is higher than a certain threshold.

In the usual literature on reputation mechanism design, the efficiency of a design is evaluated by deriving upper or lower bounds on the equilibrium payoff. One considers bounds instead of actual equilibria because it is not always obvious which of the several existing equilibria is chosen. In the following, we assume that—given truthful feedback—the seller plays the equilibrium that maximizes his payoff. This is weakly justified as some of the low payoff equilibria are unreasonable: once a buyer is asked for her feedback report, she is already in the subgame following payment to the seller (compare again Figure 1). This means that her belief of the subsequent seller action must be sufficiently high as otherwise "pay" would not have been a best response. However, it remains future work to study both upper and lower bounds on the seller strategies conditional on the subgame following "pay" being reached.

Instead of assuming honest feedback and studying design parameters, we take the complementary view and assume that we are given a sanctioning reputation mechanism which induces a certain level of seller cooperation under the assumption of honest feedback. That is, the mechanism knows the seller's strategy given honest feedback by the buyers. The feedback mechanism which we develop in this paper can then be used by a sanctioning mechanism similar to a plug-in in order to incorporate honest feedback.

### 2.3 COMMITMENT TYPES

Our game-theoretic model of the online auction setting is that of Dellarocas (2005), extended by the possibility that players can be of cooperative "commitment" types. Commitment types have a long history in the game-theoretic literature on reputation building: for example, Kreps and Wilson (1982) showed that—if the long-run players' actions are perfectly observed—a small prior belief that the long-run player is committed to cooperation is sufficient to induce a "strategic" long-run player to cooperate, as well. The intuition behind this result is that a strategic player "masks" as a commitment player, i. e. builds a reputation for being cooperative. After a sufficiently long history of cooperative play, the short-run players' belief that the long-run player is of a commitment type approaches 1 and they play a best response to cooperation. Fudenberg and Levine (1992) find a similar result for settings where the long-run players' actions are only imperfectly observed as it is the case in the online auction game. Contrasting the perfectly observable case, however, a strategic long-run player will occasionally cheat as the noise prevents the immediate revelation of his strategic nature. Cripps, Mailath and Samuelson (2004) show that in the long run these occasional plays of the cheating action eventually reveal the long-run player's strategic nature almost surely.

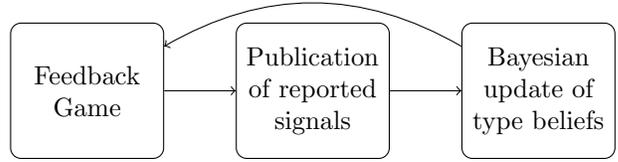

Figure 2: The embedding of the Feedback Game.

### 2.4 THE FEEDBACK GAME

Similar to the paper of Dellarocas in which he studies how often a feedback profile should be updated (2006), we subdivide the possibly infinite game into a sequence of games (henceforth, Feedback Games) with only two buyers. While these less frequent updates can generally lead to higher cooperation of strategic sellers, it is crucial to note that it cannot prevent occasional cheats by the seller in settings with imperfect monitoring of outcomes.

The Feedback Game is an extensive game with both incomplete and imperfect information. The set of players is

$$N = \{\text{seller}, 1, 2\} \quad (1)$$

which denotes the seller, the first and the second buyer. The game begins with a move by "nature" that chooses the seller's type $\theta$, which is either "strategic" or "commitment":

$$\theta \in \Theta = \{\theta_s, \theta_c\}. \quad (2)$$

Note that this move by "nature" is not an actual move but used to model the buyers' beliefs regarding the seller's type. That is, all agents share a common prior belief $Pr(\theta_c)$ that the seller is of the commitment type $\theta_c$ with $Pr(\theta_c) \geq 0$ and

$$Pr(\theta_s) + Pr(\theta_c) = 1. \quad (3)$$

Once determined, the seller learns his own type which stays fixed for the remainder of the game.

Following this move by nature, the seller decides on the quality effort level for the first buyer which is followed by a noisy signal which the buyer privately observes. She is then asked to report a signal to the mechanism. Thereafter, this procedure is repeated for the second buyer. Once both reports are elicited, they can be published. The buyers of the following Feedback Game then hold a different prior belief with regard to the seller's type. We refer to Section 3 for an explanations of their computation through Bayesian updating. See Figure 2 for the embedding of the Feedback Game.

There are $M$ levels of seller effort. To denote the exerted quality effort for buyer $i$, we use the notation

$$q^i \in Q = \{q_1, \ldots, q_M\} \quad (4)$$

with $q_M$ corresponding to full cooperation. High effort is costlier for the seller than low effort.

Let $s^i$ denote the signal received by buyer $i$

$$s^i \in S = \{s_1, s_2\} \quad (5)$$

corresponding to a low and high signal, respectively. Please note that we often write $l$ and $h$ instead of $s_1$ and $s_2$. The seller action influences buyer $i$'s signal in that cooperation makes it more probable that the signal is high. Nevertheless, it is perfectly possible that a high effort level results in a low signal. For example, a seller may send the good as described, i.e. he cooperates, but the package gets lost in the mail. It is important to note that the seller cannot observe the outcome of his action, e.g., whether the package was lost. However, all players know the probability for a certain signal given a certain effort level:

$$0 < f(h|\,q_m) < 1 \text{ for all } q_m \in Q. \quad (6)$$

Note that $f(\cdot|\,q_m)$ is a probability distribution, so that $f(l|\,q_m) = 1 - f(h|q_m)$. Also, the noise parameters are fully mixed, so that every signal has a probability greater than 0 following any quality effort. Furthermore, higher effort makes a high signal strictly more likely:

$$f(h|\,q_m) > f(h|\,q_{m-1}) \text{ for all } m = 2, \ldots, M. \quad (7)$$

Observe that the information that the seller has about the state of the game does not change between his first and his second move. He cannot observe the outcome of his first move nor the buyer's feedback report which follows it. Therefore, coalescing his two moves into a single one which combines the two original moves, results in a strategically equivalent game. Henceforth, we refer to this version of the game as the Feedback Game and denote the seller actions for the first and second buyer by

$$q^{12} = q^1 q^2 \quad (8)$$

Sellers of the commitment type always play $q_M$. As mentioned earlier, we also know a strategic seller's strategies *under the assumption of truthful buyer feedback*. We denote these by

$$\bar{q}^{12} = \bar{q}^1 \bar{q}^2 \quad (9)$$

and refer to the probability that $q_m$ is played for buyer $i$ with $Pr(\bar{q}_m^i)$. As mentioned earlier, Fudenberg and Levine (1992) show that in settings with imperfectly observed actions, the long-run player cheats occasionally so that $Pr(\bar{q}_M^i) < 1$ for $i = 1, 2$.

To incorporate external benefits from lying and participation costs, let $\Delta^i(h)$ be the external benefit buyer $i$ could gain by falsely announcing signal $h$ instead of signal $l$, the one actually received ($\Delta^i(l)$ analogously). We assume upper bounds

$$\Delta(s_d) = \max_i \Delta^i(s_d) \text{ for } s_d = l, h \quad (10)$$

on the individual external lying benefits. Furthermore, let $C^i$ be the costs reflecting buyer $i$'s time and effort required for the rating process. Similar to the lying benefits, we assume an upper bound

$$C = \max_i C^i \quad (11)$$

on the individual participation costs.

## 3 MECHANISM FRAMEWORK

In order to elicit truthful feedback and induce seller cooperation, we allow the mechanism to pay the buyers for their feedback. Before we get to the description of the payment scheme, we clarify the applied solution concept and show how to compute the conditional beliefs that the payment scheme requires.

### 3.1 SOLUTION CONCEPT

In order to have a unifying solution concept for both the seller (who shall cooperate) and the buyers (who shall be truthful), we include the seller in the definition of truthfulness. We are aware that there are other concepts that capture both information revelation and robustness to rational manipulation. However, when the monitoring of the outcomes is imperfect, it is not possible to achieve full seller cooperation, and thus robustness to rational manipulation, even if feedback is truthful. It is important to distinguish between the objectives of the sanctioning reputation mechanism and the feedback mechanism: the *sanctioning mechanism* (which we are given) induces some degree of seller cooperation if feedback is truthful. The objective of the *feedback mechanism* is to provide the sanctioning mechanism with truthful buyer feedback. That is, in noisy settings, the sanctioning mechanism can only induce *partial* cooperation given truthful feedback while the feedback mechanism can potentially elicit *perfectly* truthful buyer feedback. Since faithfulness would imply that the seller is fully cooperative in equilibrium, it is not suited for our purpose and we side to include the seller in the definition of truthfulness, instead.

**Definition 1** (Truthfulness). An equilibrium is *truthful* if and only if the seller plays $\bar{q}^{12}$ and both buyers report their signal outcomes honestly.

The equilibrium concept we apply is perfect Bayesian equilibrium (PBE) as it generalizes both Nash and subgame perfect equilibrium to extensive games with incomplete information. If, in equilibrium, there were information sets that are not reached, sequential equilibrium would further restrict beliefs on these information sets. However, the seller's sole information set is a singleton, i.e., at the time of his play, the seller has no uncertainty about the state of the world, and since $f(\cdot|\cdot)$ is fully mixed, every information set belonging to a buyer is reached with positive probability. Put differently, no buyer is ever "surprised" by an information set she encounters due to another player's deviation from equilibrium play. This has two other important implications: firstly, the best response conditions for truthfulness are not only necessary but sufficient for a PBE as the signal a buyer receives is interpreted as equilibrium play and, secondly, the Bayesian update of the type prior to the type posterior is always possible.

### 3.2 BELIEF COMPUTATIONS

One way to check whether a strategy profile is a truthful PBE is the following: pick out each player, fix the remaining players' strategies to truthfulness and check whether truthful play by the picked-out player is a best response. If the picked-out player is the seller, we have the situation of truthful feedback for which we know that the seller will play $\bar{q}^{12}$.

If the picked-out player is one of the buyers, the situation is more difficult. Without loss of generality, let buyer $i$ be the picked-out buyer and let us denote the other buyer as $r(i)$. For the best response condition to hold, it must be true that—given the seller plays $\bar{q}^{12}$ and buyer $r(i)$ is truthful—buyer $i$ is also truthful. To achieve this, we allow the mechanism to pay buyer $i$ for her feedback and condition the payment on the feedback report of $r(i)$. For pure signaling mechanisms (without a strategic seller), this so-called "peer prediction method" by Miller, Resnick and Zeckhauser (2005) has been used extensively to derive a fair number of mechanisms (Jurca and Faltings 2006; 2009, e.g.). To our knowledge, however, we are the first to study its application to sanctioning mechanisms. See Figure 3 for the procedure of a single Feedback Game.

For the moment, we assume that buyer $i$ has an incentive to report a signal. Let

$$a^i = (a_1^i, a_2^i) \qquad (12)$$

be her reporting strategy, such that she reports signal $a_j^i \in S$ if she received $s_j$. The honest strategy, i.e. always reporting the signal she received, is:

$$\bar{a}^i = (s_1, s_2) = (l, h). \qquad (13)$$

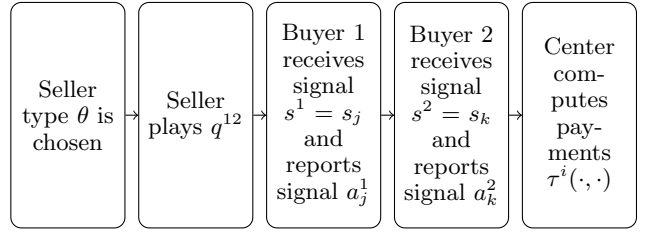

Figure 3: The feedback procedure for a single Feedback Game.

After buyer $i$ has reported a signal to the mechanism, it is compared to the report of buyer $r(i)$. Let $s^{r(i)} = s_k$ and $s^i = s_j$ denote the signals received by buyer $r(i)$ and buyer $i$, respectively. For her report, she receives a payment

$$\tau^i(a_j^i, a_k^{r(i)}) \qquad (14)$$

that depends on both her own report $a_j^i$ and the report of the other buyer $a_k^{r(i)}$. We describe how these payments are computed in Section 4.

The central idea behind the comparison of two signal reports is that knowing one of the received signals should tell you something about the other. The probability that buyer $r(i)$ received $s_k$ given $i$ received $s_j$ is defined as:

$$g^i(s_k|s_j) = Pr(s^{r(i)} = s_k|s^i = s_j). \qquad (15)$$

After receiving her signal, buyer $i$'s expected payment is thus given by Equation 16.

$$E(a_j^i = s_d|s^i = s_j) = \sum_{k=1}^{2} g^i(s_k|s_j) \tau^i(s_d, s_k). \qquad (16)$$

In the following equations, we slightly abuse the notation and write $q_m^i$ for $q^i = q_m$, i.e. the event that the seller exerts effort $q_m$ for buyer $i$ (analogously for buyer $r(i)$). We also write $\theta_c$ instead of $\theta = \theta_c$ (analogously for $\theta_s$). The signal posterior $g^i(s_k|s_j)$ can be computed with Equation 17:

$$g^i(s_k|s_j) = \sum_{m=1}^{M} f(s_k|q_m) \cdot Pr(q_m^{r(i)}|s^i = s_j). \qquad (17)$$

The probability of a specific seller effort level played for buyer $r(i)$ given the signal perceived by buyer $i$ is obtained by Bayes' law:

$$Pr(q_m^{r(i)}|s^i = s_j) = \frac{Pr(s^i = s_j|q_m^{r(i)}) \cdot Pr(q_m^{r(i)})}{Pr(s^i = s_j)}. \qquad (18)$$

The prior signal probability for buyer $i$ is:

$$Pr(s^i = s_j) = \sum_{m=1}^{M} f(s_j|q_m) \cdot Pr(q_m^i). \qquad (19)$$

The prior probability for a specific seller effort level played for buyer $i$ which is required for Equations 18, 19 and 24 can be computed with Equation 20:

$$Pr(q_m^i) = Pr(q_m^i|\theta_c) \cdot Pr(\theta_c) \\ + Pr(q_m^i|\theta_s) \cdot Pr(\theta_s). \quad (20)$$

The values for $Pr(q_m^i|\theta)$ are derived with either of the two following equations:

$$Pr(q_m^i|\theta_c) = \begin{cases} 1, & \text{if } m = M \\ 0, & \text{else} \end{cases} \quad (21)$$

and

$$Pr(q_m^i|\theta_s) = Pr(\bar{q}_m^i). \quad (22)$$

For Equation 18, we require $Pr(s^i = s_j | q_m^{r(i)})$. At first sight, it may seem that the probability of the seller effort level for buyer $r(i)$ is independent of the other buyer's signal. However, the materialized $q^{r(i)}$ tells us something about the seller's type which, in turn, influences buyer $i$'s signal:

$$Pr(s^i = s_j | q_m^{r(i)}) = Pr(s^i = s_j | \theta_c) \cdot Pr(\theta_c | q_m^{r(i)}) \\ + Pr(s^i = s_j | \theta_s) \cdot Pr(\theta_s | q_m^{r(i)}). \quad (23)$$

The type probability knowing the seller's effort level for a certain buyer is again computed by Bayes' law:

$$Pr(\theta | q_m^{r(i)}) = \frac{Pr(q_m^{r(i)}|\theta) \cdot Pr(\theta)}{Pr(q_m^{r(i)})}. \quad (24)$$

Finally, Equation 25 shows how to obtain the probability for a signal knowing the seller's type.

$$Pr(s^i = s_j | \theta) = \sum_{m=1}^{M} f(s_j | q_m) \cdot Pr(q_m^i | \theta). \quad (25)$$

We can now compute the signal posterior $g^i(s_k | s_j)$ and with it the expected payment to buyer $i$ given truthfulness by both buyer $r(i)$ and the seller.

The Bayesian update for the type beliefs, once the two signals are elicited (compare Figure 2), are computed with Equation 26 (using Equations 19 and 25):

$$Pr(\theta | s^i = s_j) = \frac{Pr(s^i = s_j | \theta) \cdot Pr(\theta)}{Pr(s^i = s_j)}. \quad (26)$$

It is easy to see that with a fully mixed $f(\cdot|\cdot)$, and a positive prior probability for both types in the first Feedback Game, both type probabilities stay positive for every Feedback Game that follows.

# 4 THE PAYMENT SCHEME

We follow Jurca and Faltings (2006) and formulate the payment scheme as a Linear Program (LP) that minimizes the required budget. The technique of solving a mechanism design problem using mathematical optimization is called Automated Mechanism Design and was first introduced by Conitzer and Sandholm (2002).

## 4.1 THE LINEAR PROGRAM

The constraints of the LP are divided into two groups. The first group consists of the honesty constraints which require that an honest signal announcement by buyer $i$ is the single best response to a truthful report by $r(i)$ and a truthful seller. For every possible signal observation $s^i = s_j \in S$, there exists a dishonest announcement $a_j^i \neq \bar{a}_j$. We want the expected payment of a truthful report by agent $i$ to be larger than the expected payment of the other possible report. More accurately, incorporating external lying incentives, we want it to be larger by a margin greater than the respective $\Delta(\cdot)$. Note that the buyer's valuation for a signal can be left out as the *received* signal is independent of the signal *announcement*:

$$\sum_{k=1}^{2} g^i(s_k | s_j) \left( \tau^i(s_j, s_k) - \tau^i(s_d, s_k) \right) > \Delta(s_d). \quad (27)$$
$$\forall s_j, s_d \in S, s_j \neq s_d$$

The second group consists of the participation or individual rationality (IR) constraints. A rational buyer will only give feedback if she is remunerated with at least as much as the rating process costs her. In order to avoid indifference between participation and absence we demand that, at the time of her participation decision, the agent receives an expected payment that is *higher* than $C$. Since the buyers decide whether to participate only after they have experienced the product, i.e. after they received their signal, participation needs to be better than non-participation given any of the two possible signal observations (*interim* IR):

$$\sum_{k=1}^{2} g^i(s_k | s_j) \cdot \tau^i(s_j, s_k) > C, \quad \forall s_j \in S. \quad (28)$$

As mentioned earlier, the objective of the mechanism is to minimize its required budget. Please note that technically we can only achieve a very good approximation of the required budget as we have strict inequalities in our constraints. The budget $B$ is the expected payment given a certain signal weighted with the signal's prior probability. Together with the assumption that there is no possibility to withdraw credit from the agents, so that all payments are non-negative, the summarized payment scheme formulated as an LP is LP 1. Note that we have to add a small $\epsilon > 0$ to the right side of both the honesty and participation constraints as the definition of Linear Programs does not include strict inequalities.

**LP 1.**

$$\min. \ B = \sum_{j=1}^{2} Pr(s_j) \left( \sum_{k=1}^{2} g^i(s_k|\, s_j) \cdot \tau^i(s_j, s_k) \right)$$

$$s.t. \ \sum_{k=1}^{2} g^i(s_k|\, s_j) \left( \tau^i(s_j, s_k) - \tau^i(s_d, s_k) \right) \geq \Delta(s_d) + \epsilon$$

$$\forall s_j, s_d \in S, s_j \neq s_d$$

$$\sum_{k=1}^{2} g^i(s_k|\, s_j) \cdot \tau^i(s_j, s_k) \geq C + \epsilon \quad \forall s_j \in S$$

$$\tau^i(s_j, s_k) \geq 0; \quad \forall s_j, s_k \in S$$

### 4.2 FEASIBILITY ANALYSIS

It is clear that LP 1 is bounded since all factors in the objective function are non-negative. To study if and when it is feasible, we firstly prove three lemmas that eventually enable us to prove our main statement that LP 1 is feasible if and only if both seller types have prior probability greater 0. The intuition for this result is that a buyer's belief about another buyer's experience does not change if there is only a single seller type whose strategies are commonly held a-priori.

**Lemma 1.** *LP 1 is feasible if and only if $g^i(h|\, h) \neq g^i(h|\, l)$ for $i = 1, 2$.*

*Proof.* For reasons of clarity, we omit the letter $i$. We first reduce the feasibility of LP 1 to the feasibility of its honesty constraints with no external lying benefits:

**FP 1.**

$$\sum_{k=1}^{2} g(s_k|\, s_j) \cdot \tau(s_j, s_k) - \sum_{k=1}^{2} g(s_k|\, s_j) \cdot \tau(s_d, s_k) \geq \epsilon$$

$$\forall s_j, s_d \in S, s_j \neq s_d$$

We proceed by transforming a feasible solution of FP 1 into a feasible solution of LP 1 and begin with incorporating $\Delta(s_d)$. Let $\epsilon'$ denote the maximal of all possible $\epsilon$ which corresponds to FP 1's "honesty margin". As expected utility is invariant to affine transformations, we can multiply $\tau(\cdot, \cdot)$ with a constant factor $\gamma$ without changing the incentive properties. Let $\Delta' = \max_d \Delta(s_d)$ denote the maximal external lying benefit. We choose $\gamma = \frac{\Delta'}{\epsilon'}$, so that after multiplying $\tau(\cdot, \cdot)$ with $\gamma$, the resulting feasibility program incorporates all external lying benefits.

If the minimal $\tau(s_j, s_k)$ is negative, let $\tau'$ denote its absolute value. Incorporating participation costs $C$ and ensuring that all $\tau(s_j, s_k)$ are positive is done by adding $\tau' + C + \epsilon$ to all $\tau(s_j, s_k)$. As mentioned, expected utility is invariant to affine transformations, so that adding a constant does not change the agents' incentives.

We change the right side of the inequality from "$\geq \epsilon$" to "$> 0$", so that the two honesty constraints can be written as:

$$g(h|\, h) \underbrace{(\tau(h,h) - \tau(l,h))}_{:=\tau_1} + g(l|\, h) \underbrace{(\tau(h,l) - \tau(l,l))}_{:=\tau_2} > 0$$

$$g(l|\, l) \underbrace{(\tau(l,l) - \tau(h,l))}_{-\tau_2} + g(h|\, l) \underbrace{(\tau(l,h) - \tau(h,h))}_{-\tau_1} > 0.$$

Rewriting $g(l|\, h) = 1 - g(h|\, h)$ and $g(l|\, l) = 1 - g(h|\, l)$ and multiplying the second line by $-1$, we obtain:

$$g(h|\, h)\tau_1 + \tau_2 - g(h|\, h)\tau_2 > 0$$
$$g(h|\, l)\tau_1 + \tau_2 - g(h|\, l)\tau_2 < 0.$$

It is obvious that if $g(h|\, h) = g(h|\, l)$, no assignment that satisfies both constraints can be found. If $g(h|\, h) > g(h|\, l)$, pick a value $g'$ so that $g(h|\, h) > g' > g(h|\, l)$ and assign $\tau_1 = 1 - g'$ and $\tau_2 = -g'$. We show that this assignment satisfies both constraints:

$$g(h|\, h)(1 - g') - g' + g(h|\, h) \cdot g'$$
$$= g(h|\, h) - g(h|\, h) \cdot g' - g' + g(h|\, h) \cdot g'$$
$$= g(h|\, h) - g' > 0$$

and

$$g(h|\, l)(1 - g') - g' + g(h|\, l) \cdot g'$$
$$= g(h|\, l) - g' < 0$$

The case $g(h|\, l) > g' > g(h|\, h)$ is analogous with $\tau_1 = g' - 1$ and $\tau_2 = g'$. □

We require two other lemmas for the proposition.

**Lemma 2.** $Pr(s^i = h|\theta_c) > Pr(s^i = h) > Pr(s^i = h|\theta_s)$ *if* $0 < Pr(\theta_c) < 1$.

*Proof.* Sellers of the commitment type always play full cooperation, so that $Pr(s^i = h|\theta_c) = f(h|\, q_M)$. Sellers of the strategic type play $Pr(s^i = h|\theta_s) = \sum_{m=1}^{M} f(h|q_m) \cdot Pr(\bar{q}_m^i)$ (Equations 22 and 25). It then follows from $f(h|\, q_M) > f(h|\, q_{M-1})$ and $Pr(\bar{q}_M^i) < 1$ that $Pr(s^i = h|\theta_c) > Pr(s^i = h|\theta_s)$. The statement then follows from $Pr(s^i = h) = Pr(s^i = h|\theta_s) \cdot Pr(\theta_s) + Pr(s^i = h|\theta_c) \cdot Pr(\theta_c)$. □

Using Lemma 2 we can then prove Lemma 3:

**Lemma 3.** $Pr(\theta_c|\, s^i = h) > Pr(\theta_c|\, s^i = l)$ *for* $i = 1, 2$ *if* $0 < Pr(\theta_c) < 1$.

*Proof.*

$$Pr(\theta_c|\, s^i = h) = Pr(\theta_c) \cdot \frac{Pr(s^i = h|\theta_c)}{Pr(s^i = h)}$$
$$> Pr(\theta_c) \cdot \frac{Pr(s^i = h)}{Pr(s^i = h)}$$
$$= Pr(\theta_c)$$
$$> Pr(\theta_c) \cdot \frac{Pr(s^i = l|\theta_c)}{Pr(s^i = l)}$$
$$= Pr(\theta_c|\, s^i = l)$$

□

We can now prove the following proposition.

**Proposition 4.** *LP 1 is feasible if and only if $0 < Pr(\theta_c) < 1$.*

*Proof.* First, we show that from either $Pr(\theta_c) = 0$ or $Pr(\theta_c) = 1$, it follows that $g(h|h) = g(h|l)$. Thereafter, we show that assuming both $g(h|h) = g(h|l)$ and $0 < Pr(\theta_c) < 1$ leads to a contradiction.

If $Pr(\theta_c) = 0$, the signal received by buyer $r(i)$ is simply $Pr(s^{r(i)} = s_k|\theta_s) = \sum_{m=1}^{M} f(s_k|q_m) \cdot Pr(\bar{q}_m^{r(i)})$ and thus independent of the signal received by buyer $i$ and, in particular, it thus holds that $g(h|h) = g(h|l)$. Similarly, $Pr(\theta_c) = 1$ implies that buyer $r(i)$ receives a high signal with probability $f(h|q_M)$ which is also independent of the signal received by buyer $i$.

Let us then assume that both $g(h|h) = g(h|l)$ and $0 < Pr(\theta_c) < 1$ holds. These are contradictory:

$$\begin{aligned}
g(h|h) =\ & Pr(s^{r(i)} = h|\theta_c) \cdot Pr(\theta_c|s^i = h) \\
& + Pr(s^{r(i)} = h|\theta_s) \cdot Pr(\theta_s|s^i = h) \\
=\ & f(h|q_M) \cdot Pr(\theta_c|s^i = h) + Pr(s^{r(i)} = h|\theta_s) \\
& - Pr(s^{r(i)} = h|\theta_s) \cdot Pr(\theta_c|s^i = h) \\
=\ & Pr(\theta_c|s^i = h) \left( f(h|q_M) - Pr(s^{r(i)} = h|\theta_s) \right) \\
& + Pr(s^{r(i)} = h|\theta_s) \\
>\ & Pr(\theta_c|s^i = l) \left( f(h|q_M) - Pr(s^{r(i)} = h|\theta_s) \right) \\
& + Pr(s^{r(i)} = h|\theta_s) \\
=\ & f(h|q_M) \cdot Pr(\theta_c|s^i = l) + Pr(s^{r(i)} = h|\theta_s) \\
& - Pr(s^{r(i)} = h|\theta_s) \cdot Pr(\theta_c|s^i = l) \\
=\ & Pr(s^{r(i)} = h|\theta_c) \cdot Pr(\theta_c|s^i = l) \\
& + Pr(s^{r(i)} = h|\theta_s) \cdot Pr(\theta_s|s^i = l) = g(h|l) \notmid
\end{aligned}$$

□

**Corollary 5.** *If both seller types have prior probability greater than 0 and every buyer is paid for her report according to payments computed by LP 1, truthfulness is a perfect Bayesian equilibrium in the Feedback Game.*

If LP 1 is feasible and bounded, the proof follows immediately from the design of the payment scheme.

## 5 CONCLUSION

We have shown that the peer prediction method can be used to design a truthful feedback scheme for sanctioning mechanisms. Drawing on the literature on game theory and reputation building, we introduced a commitment type seller who is fully cooperative and showed that its presence is critical for the method's applicability.

In future work, we will reduce the common knowledge assumptions of our mechanism. In particular, we will modify the payment scheme to allow for bounds on the probabilistic parameters instead of the (possibly unknown) specific numbers. For example, it should be sufficient to know that high seller effort is followed by a high signal with probability of *at least 90%*. Another difficulty with regard to the mechanism's application is the assumption of a long-lived seller. We will thus study mechanisms that no longer rely on a long history of past feedback reports but instead pay a seller an amount directly dependent on the reported quality and only after the transaction.

## Acknowledgements

We thank Malte Helmert, Sven Seuken, Boi Faltings and three anonymous reviewers for helpful comments. The author is partially financed by a Fellowship from the *Landesgraduiertenförderung Baden-Württemberg*.